\documentclass[preprint,12pt]{modified-elsarticle}

\usepackage{amssymb}

\usepackage[version=4]{mhchem}

\usepackage{url}

\usepackage{subcaption}

\usepackage{array}

\def\today{November 30, 2015}

\begin{document}

\begin{frontmatter}



\title{Band structure and transport studies of half Heusler compound DyPdBi: An efficient thermoelectric material}


\author[1] {S. Krishnaveni}
\ead{sarathyveni@gmail.com}

\author[1] {M. Sundareswari}
\ead{sundare65@gmail.com}

\author[2] {P. C. Deshmukh}
\ead{pcd@physics.iitm.ac.in}

\author[3,4] {S. R. Valluri}
\ead{valluri@uwo.ca, vallurisr@gmail.com}

\author[3] {Ken Roberts}
\ead{krobe8@uwo.ca}

\address[1]{Department of Physics, Sathyabama University, Chennai, 600119, India}

\address[2]{Department of Physics, Indian Institute of Technology Madras, Chennai, 600036, India}

\address[3]{Department of Physics and Astronomy, The University of Western Ontario, London, N6A 3K7, Canada}

\address[4]{Kings University College, Department of Economics, Business and Mathematics, London, Canada}

\begin{abstract}
The discovery of Heusler alloys has revolutionized the research field
 of intermetallics due to the ease with which one can derive potential
 candidates for multifunctional applications.
During recent years, many half Heusler alloys have been investigated
 for their thermoelectric properties.
 The f electron based rare earth ternary half Heusler compound \ce{DyPdBi}
 has its f energy levels located close to the Fermi energy level.
Other research efforts have emphasized that such materials
 have good thermoelectric capabilities.
We have explored using first principles
 the electronic band structure of \ce{DyPdBi}
 by use of different exchange correlation potentials
 in the density functional theoretical framework.
Transport coefficients that arise in
the study of thermoelectric properties of \ce{DyPdBi}
 have been calculated and illustrate its potential
 as an efficient thermoelectric material.
Both the theoretically estimated Seebeck coefficient
 and the power factor
 agree well with the available experimental results.
Our calculations illustrate that it is essential
 to include spin-orbit coupling
 in these models of f electron half Heusler materials.
\end{abstract}


\end{frontmatter}



\section{Introduction}
\label{sect-intro}
Intermetallic compounds play a vital role in designing new
 materials for various technology applications including
 refractory and structural applications, spintronics and
 magnetic and superconductivity applications.
Ab-initio study on rhodium based intermetallic compounds
 \ce{Rh3X} (X= Ti, Zr, Hf, V, Nb or Ta)
 has been carried out by
 Rajagopalan, et al \cite{raja-2004} at ambient conditions
 and under compression.
 The discovery of Heusler alloys \cite{heusler-1903}
 has revolutionized the research field on intermetallics,
 since one can derive potential candidates for
 multifunctional applications.
 Full Heusler alloys (\ce{R2TX} type where R= rare-earth,
 T= transition metal and X= p-electron element of the
 periodic table) with $L2_{1}$ structure,
 and half Heusler alloys (\ce{RTX} type) with $C1_b$ structure,
 have been studied
 \cite{kubler-1983, raphael-2002, cheng-2001} mainly for
 their half metallic character.
Many such materials are reported to be ferromagnetic;
 some are spintronic with 100\% spin polarization.

Several half Heusler alloys
 \cite{wurmehl-2005, galanakis-2006, svetl-2006, nanda-2003,
 nanda-2005, muta-2006, larson-1999, jesus-2014, downie-2014,
 gofryk-2011, nakajima-2015, sawai-2010, bose-2011, pan-2013,
 galana-2002}
 have been prepared and reported by experimentalists
 and also by theoreticians.
Gillessen \cite{gille-2009, drons-2009, drons-2010}
 has theoretically predicted and reported many full,
 and half, Heusler compounds using Density Functional Theory.
Band structure calculations on one such Heusler alloy,
 namely \ce{Ir2CrAl}, have been reported by
 Krishnaveni, et al \cite{krishna-2015} which reveal
 100\% spin polarization, half-metallicity,
 and ferromagnetism, with a magnetic moment of $3\mu_B$.
The properties observed in half Heusler compounds include
 half metallicity, semiconductivity, giant magnetoresistance,
 or heavy Fermion state.

Increasing demand for energy necessitates finding
 alternative ways of generating electricity.
Thermoelectric materials generate electric current using
 the Seebeck effect.
The major challenge is the generation of very high
 thermoelectric current.
Over the years many half Heusler alloys have been
 investigated for their thermoelectric properties.
The figure of merit (ZT) value characterizes the efficiency
 of the thermoelectric property of the given material.
$ZT= \sigma S^{2}T/\kappa$
 where $T$ is the temperature,
 $\sigma$ is the electrical conductivity,
 $S$ is the Seebeck coefficient
 and $\kappa$ is the thermal conductivity which contains
 both the electronic and the lattice contributions.
Increased efficiency of thermoelectric performance
 of a given material can be sought by enhancing its
 Seebeck coefficient, its electrical conductivity
 or lowering its electronic/lattice thermal conductivity.
That can be attained by tuning the electronic structure
of the given material \cite{muechler-2013}.

Mahan and Sofo \cite{mahan-1996} have reported
 that if the f-levels of any given system
 lie close to the Fermi level,
 then that system has good thermoelectric response.
That motivates us to consider an f-electron based
 rare earth (R) ternary half Heusler compound,
 namely \ce{DyPdBi},
 which has been synthesized and reported earlier
 \cite{gofryk-2011}.
This compound \ce{DyPdBi} crystallizes \cite{gofryk-2011}
 in the cubic structure of \ce{MgAgAs} type,
 with the space group 216 (F-43m)
 where \ce{Dy} occupies the position (0, 0, 0),
 \ce{Pd} occupies position (3/4, 3/4, 3/4),
 and \ce{Bi} occupies the (1/2, 1/2, 1/2) position
 \cite{larson-1999}.
Experimental investigations \cite{gofryk-2011}
 show that \ce{DyPdBi} is antiferromagnetic up to 3.5K,
 exhibits metallic behaviour at low temperature
 and is semiconducting at high temperature.
	
In this paper, we report the electronic and transport
 properties of the \ce{DyPdBi} compound as predicted
 by first principles electronic band structure calculations.
Although much theoretical work has been reported
 for quite a number of thermoelectric materials,
 f-electron systems have not been explored extensively
 to the best of our knowledge.
As one has to compare and look for possible accord
 between the existing experimental results for \ce{DyPdBi}
 with that of the theory of
 strongly correlated f-electron systems,
 to begin with we performed band structure calculations
 on \ce{DyPdBi} by applying different
 exchange correlation potentials.
In order to study the thermoelectric properties
 of \ce{DyPdBi}, the transport coefficients have been
 calculated by using Boltzmann theory.
The BoltzTraP code,
 developed by Madsen and Singh \cite{madsen-2006}
 as implemented in the Wien2k package,
 was used to calculate the Seebeck coefficient,
 power factor, and other transport coefficients.

\section{Computational Details}
\label{sect-comput}

First principles band structure calculations have been
 performed on the \ce{DyPdBi} compound using the
 full potential linear augmented plane wave (FP-LAPW)
 method with PBEsol-GGA (Perdew etal-08) exchange
 correlation as implemented in the Wien2k (13.1) code
 \cite{blaha-2014} by generating the structure file
 with an experimental lattice
 parameter \cite{gofryk-2011} of $a_{exp}=12.5388$ a.u.
The FP-LAPW method includes a procedure for solving
 Kohn-Sham equations for the ground state density,
 total energy and eigenvalues of a many electron system
 by introducing a basis set.
In order to define the basis set,
 the given unit cell was divided into non-overlapping,
  touching spheres called muffin-tin (MT) spheres.
The muffin-tin potential is assumed to be
 spherically symmetric close to the nuclei/core and is
 assumed to be plane in the interstitial region.
The solutions of the Schr\"odinger equation outside
 the MT spheres are plane waves;
 in this region, the potential is constant.
The solutions of the Schr\"odinger equation inside the
 MT spheres are made up of angular momentum partial waves.
The solutions for the entire space are obtained by matching
 the two solutions at the boundaries of the MT spheres.
The convergence of the combined basis set is controlled
 by a cutoff parameter $R_{MT}*K_{max}$,
 where $R_{MT}$ is the smallest muffin-tin sphere radius
 in the unit cell and
 $K_{max}$ is the magnitude of the largest K-vector.
Also, the charge leakage from the core is kept minimal.
The radii of the muffin-tins employed were 2.6 a.u.,
 2.6 a.u. and 2.62 a.u. \cite{cottenier-2002}
 respectively for \ce{Dy}, \ce{Pd} and \ce{Bi},
 for the calculations performed initially
 without spin orbit coupling (SOC)
 on a k-mesh of 31x31x31 where
 $RK_{max}=7$ and $G_{max}=10$ \cite{yang-2008}.
Volume optimisation and fitting in equation of states
 gives the optimised lattice constant as
 $a_{opt}=12.5519$ a.u.
	
In general, the electron's intrinsic spin couples
 with its orbital spin angular momentum
  and give rise to the total angular momentum.
All energy levels except for the s- states
 of one electron atoms are thus split into two sub-states.
This produces the fine structure of the spectral lines.
The spin-orbit interaction scales as $Z^{2}$,
 hence it is important for atoms with high atomic numbers.
For the present f-electron based \ce{DyPdBi} compound,
 one has to consider the effect of spin orbit coupling.
Accordingly, the above calculations were
 repeated by including SOC.
In both the cases,
 self consistent field calculations were performed,
  with an energy convergence of 0.0001 Ryd,
  and charge convergence of 0.0001e.
Splitting of f-electron bands of dysprosium
 at the Fermi level is well demonstrated by
 the Density of States (DoS) histograms
 that were drawn with SOC (see figure 2).
This result ensures the good thermoelectric response
 of \ce{DyPdBi}
 as suggested by G.D.Mahan and J.O. Sofo \cite{mahan-1996}.
	
In order to explore the band structure of the
 f-electron based ternary half Heusler compound \ce{DyPdBi},
 the band structure calculation has been extended
 by performing spin polarized calculations
 (as \ce{DyPdBi} is experimentally reported to be magnetic)
 with Local Spin Density Approximation (LSDA),
 and with the Perdew-Burke-Ernzerhof (PBE)
 exchange correlation potential.
The spin polarized calculation was performed with
 a k-mesh of 17x17x17.
$RK_{max}=7$ and $G_{max}=14$ were used.
The radii of the muffin-tin orbitals were
 2.5 a.u. for \ce{Dy}, 2.55 a.u. for \ce{Pd},
 and 2.6 a.u. for \ce{Bi}.
In this case also, self consistent calculations were carried
 out with the energy convergence of 0.0001 Ry
 and charge convergence of 0.0001 e.
The optimized lattice parameter was calculated for each of
 the above mentioned exchange correlation schemes by
 fitting the volume optimization into the Birch-Murunaghan
 equation of state.
The optimised lattice parameter ($a_{opt}$),
 total energy ($E_{tot}$), Fermi energy($E_{F}$),
 bulk modulus ($B$),
 density of states at Fermi level (N($E_{F}$))
 of both the spin-up and spin-down states and
 spin magnetic moment/unit cell ($\mu_{B}$)
 values of \ce{DyPdBi} compound,
 by LSDA and PBE correlation schemes,
 are presented in table 1.
The present compound \ce{DyPdBi} is comprised of
dysprosium (\ce{Dy}) which has half filled f orbitals
 and is strongly correlated.
The behaviour of electrons in such systems can be described
 well by including the onsite Coulomb interaction
 in the conventional Hamiltonian.
Hence the calculations were done by including the
 Hubbard potential for dysprosium.
The Hubbard potential explicitly includes the Coulomb
 parameter ($U$) and Coulomb exchange parameter $J$
 for the 4f electrons.
The value of the Hubbard potential for dysprosium is
 based upon the literature \cite{saini1-2007}
 with U=0.49 Ryd and J=0.05 Ryd.
Coulomb-corrected local spin density approximation
 (LSDA + U) was applied, as this method,
 introduced by Anisimov, et al \cite{anisimov-1991},
 is found to be very efficient for several rare-earth
 compounds \cite{saini1-2006, saini2-2007, saini3-2007,
 saini2-2006, saini4-2007, saini5-2007, saini-2011}.

\section{Results and Discussion}
\label{sect-discuss}

\subsection{Structural, Electronic and Magnetic properties of \ce{DyPdBi}}

From table 1, it is clear that LSDA underestimates
 the experimental lattice parameter of \ce{DyPdBi}
 whereas the PBE calculation overestimates the same.
Due to this, there is considerable shift in the
 Fermi energy level and fluctuation in the
 bulk modulus values of the respective scheme.
The density of states at the Fermi energy level
 is very low for the spin-up states
 in comparison to that for the spin down states.
The spin magnetic moment per unit cell ranges
 between 4.73 $\mu_{B}$ and 4.87 $\mu_{B}$.

\subsubsection{Density of States (DoS)}

The total and partial density of states histograms
 of \ce{DyPdBi} as calculated for PBEsol
 without spin orbit coupling
 are shown in figures 1(a) and 1(b).
From figure 1(a), one finds that
 the total DoS of the compound \ce{DyPdBi}
 is around 110 states/eV at the Fermi energy level,
 of which around 90 states/eV
 are contributed mainly by dysprosium.
Figure 1(b) suggests that it is only the \ce{Dy}-f states
 that contribute to the total DoS of the compound.
Thus, nearly 80\% of the total DoS at Fermi level is
 due to \ce{Dy}-f states alone.

The total and partial density of states histograms of
 \ce{DyPdBi} for PBEsol with spin orbit coupling are
 presented in figures 2(a) and 2(b).
 From these figures, one can clearly see that the \ce{Dy}-f
 states split into two levels;
 one level comprised of a few states below the Fermi level
 and the other comprised of a few states at the Fermi level.
The states below the Fermi level are shifted down
 by 1.0 eV from the Fermi level.
Figure 2(b) shows that the total DoS at the Fermi level
 is nearly 54 states/eV, of which 45 states/eV
 are contributed by \ce{Dy}-f states.
Comparison of figures 1 and 2 suggest that the inclusion
 of spin orbit coupling in this calculation
 has reduced the influence of
 \ce{Dy}-f states by approximately 50\% at the Fermi level
 besides the splitting \cite{larson-1999} that is caused.
		
Though we performed spin polarised calculations
 for various exchange correlation schemes, viz.,
 Local Spin Density Approximation (LSDA),
 PBE and (LSDA + U) on \ce{DyPdBi},
 the DoS histograms are analyzed for LSDA and (LSDA + U)
 \cite{larson-1999, saini1-2006, saini2-2007, saini3-2007,
 saini2-2006, saini4-2007, saini5-2007, saini-2011}
 alone as these methods are known to be rather efficient
 for many of the rare-earth compounds.
Figures 3(a) and 3(b) show the total and partial DoS
 histograms of \ce{DyPdBi} using LSDA for spin-up
 and spin-down states respectively.
Similarly figures 4(a) and 4(b) show the same for (LSDA+U).
From figures 3(a) and 4(a), drawn for spin-up states,
 it is clear that the total, and hence the partial,
 DoS is very low of the order of 3-4 states/eV
 at the Fermi energy level.
Inclusion of the Hubbard potential is not of much
 significance at the Fermi level for spin up states.
From figures 3(b) and 4(b), drawn for spin-down states,
 it can be inferred that the total DoS at the Fermi energy
 level is almost all due to dysprosium
 and has increased from 26 states/eV to 30 states/eV.
The inclusion of the Hubbard potential did not have much
 influence on the DoS of dysprosium at the
 Fermi energy level.

\subsubsection{Band Structure}

Electronic structure plays a major role in
 understanding the properties of a given material
 under ambient conditions.
The band structure of the compound \ce{DyPdBi}
 is shown in figures 5(a) and 5(b) without SOC,
 and with SOC, respectively.
From figure 5(a), one can infer that without SOC
 the f-states of dysprosium spread around the Fermi level,
 making the compound more metallic.

With the inclusion of SOC, from figure 5(b),
 it is observed that at the $\Gamma$ point
 the f-states of dysprosium undergo splitting.
The splitting is not as significant
 for the k-points that run from L-W and X-K points.

Figures 6(a) and 6(b) represent the band structure
 of spin-up states of \ce{DyPdBi} for LSDA and (LSDA + U).
Analysis of figure 6 shows that the spin-up energy bands
 of both LSDA and (LSDA+U) have a very narrow peak
 at the $\Gamma$ point and above the Fermi level,
 and the gap between adjacent bands
 comparatively increases in the LSDA+U calculation.
This again depicts the metallic nature of \ce{DyPdBi}.
Mahan \cite{mahan-1996} has reported that
 the f-electron states which are more tightly bound
 in atoms bind little in solids and as a result
 they give larger contributions to the density of states
 which are of Lorentzian shape with rather narrow widths.
In the present case of \ce{DyPdBi} also,
 the spin-up states band structure displays such a narrow
 peak at the Fermi energy level.
The existence of such a Lorentzian narrow peak
 with a sharp singularity close to the Fermi level
 is a signature for a potential thermoelectric material.
The Lorentzian peak near the Fermi level
 enhances the electrical conductivity in spin up states.
	
In contrast, the spin-down energy bands
 of LSDA and (LSDA+U) shown in figures 7(a) and 7(b) are
 flat, and they spread just above and below the Fermi level.
Further, one can see that the conduction band minimum
 lies almost along the Fermi energy level
 except in the vicinity of the $\Gamma$ point,
 but the valence band maximum lies just below
 the Fermi level at about 0.2eV at the $\Gamma$ point,
 showing semiconducting-like nature.
The presence of such flat bands at the valence band
 maximum corresponds to a large effective mass and
 is another key factor to improve the thermoelectric
 efficiency \cite{rameshe-2015}.
Thereby it is inferred that \ce{DyPdBi} could be
 semimetallic since it exhibits metallic nature
 for spin-up band structure calculations,
 and semi-conducting like behaviour for
 spin-down calculations.

\subsection{Transport Properties}

The transport properties of \ce{DyPdBi}
 were calculated by using
 the BoltzTraP code \cite{madsen-2006}
 interfaced to the Wien2k program.
For background, we refer to the description of 
 Rameshe, et al \cite{rameshe-2015}:
The calculations are based on a
 semi-classical treatment for
 the solution of the Boltzmann equation
 utilizing 
 the relaxation time approximation and
 the rigid body approximation.
The Seebeck Coefficient ($S$),
 as a function of temperature ($T$) and doping ($\rho$),
 is given by
\begin{equation*}
  S(T,\rho) = \frac{\int dE \, \sigma(E)(E-\mu) \, df/dE}
                   {\int dE \, \sigma(E) \, df/dE}
\end{equation*}
 where $f$ is the Fermi function and
 $\mu$ is the chemical potential.
Here $\sigma(E)$ is the transport function given by
\begin{equation*}
  \sigma(E)= N(E) \, V^{2}(E) \, \tau(E)
\end{equation*}
 where $N(E)$ is the density of states,
 $V(E)$ is the band velocity and
 $\tau(E)$ is the scattering time.
Under the constant scattering time approximation
 $\tau(E)$ is independent of energy \cite{singh-2010}.
The thermoelectric power factor $S^{2}\sigma$
 is determined by the term $\sigma(T)$
 which is given by
\begin{equation*}
  \sigma (T) = \int dE \, \sigma(E) \, df/dE.
\end{equation*}

The transport properties of \ce{DyPdBi}
 based on PBEsol-GGA without SOC, with SOC
 and with (LSDA + U) calculations
 were estimated by applying the BoltzTraP code.
The values of Seebeck coefficient, power factor,
 resistivity $\rho$, $\sigma/\tau$ and
 $S^{2}\sigma/\tau$ at 300K for Fermi energy
 thus calculated are listed in table 2.
Analysis of this table shows that there is good agreement
 of our theoretical results with the available
 experimental data \cite{gofryk-2011}
 which give the Seebeck coefficient of 92 $\mu V/K$,
 power factor of 20$\mu$ Wcm$^{-1}$ K$^{-2}$
 and the resistivity of 422$\mu\Omega$ -cm.
The Boltztrap output that has been calculated
 for the scheme with SOC at 300K for Fermi energy
 gives the Seebeck coefficient value to be 138$\mu$V/Kelvin
 and the power factor to be 30.8$\mu$W$cm^{-1}K^{-2}$.
The estimated power factor value matches well
 with that of existing potential thermoelectric materials,
 namely \ce{TiCoSb} \cite{yang-2008} based alloys,
 that have a  maximum power factor of
 23 $\mu$W$cm^{-1} K^{-2}$ at 850K
 and 34 $\mu$W$cm^{-1} K^{-2}$
 for \ce{ZrNiSn} based alloys \cite{yang-2008}.
The obtained Seebeck coefficient value matches well
 with that of yet another potential TE material,
 namely \ce{Fe2VAl} \cite{nishino-2011},
 whose Seebeck coefficient is -130 $\mu$V/Kelvin at 300K.

In order to look for the experimental positive
 Seebeck coefficient value,
 graphs are drawn and are explained below.
The variation of Seebeck coefficient,
 $S^{2}\sigma/\tau$ values, with the total energy
 is plotted for PBEsol-GGA without SOC,
 with SOC, and with (LSDA + U) - spin down
 at temperatures 200K, 300K and 400K.
These plots are presented in figures 8 and 9.
In these figures, the reference level at zero
 represents the Fermi energy level in eV and
 the relaxation time is taken as
 0.8x$10^{-14}$ second as suggested in
 the BoltzTraP user manual.
Analysis of figures 8(a)-8(c) reveals a peak
 in the Seebeck coefficient value
 and it remains positive just below the Fermi energy
 for temperatures 200K, 300K and 400K.
This is in accordance with the
 reported literature \cite{gofryk-2011} that suggests
 the possibility that holes play a major role
 in the electrical and heat transport properties
 of the \ce{DyPdBi} compound.
Further, in figure 8(a), we see that the Seebeck
 coefficient is maximum at 200K and decreases with
 increases in temperature,
 whereas in figures 8(b) and 8(c),
 it remains stable with temperature.

$S^{2}\sigma/\tau$ is another important parameter
 to characterize a thermoelectric material
 as found by the research group of Dresselhaus
 and others \cite{poudel-2008, dressel-2007, hicks-1997}.
Its variation with energy for \ce{DyPdBi}
 at temperatures 200K, 300K and 400K
 is shown in figures 9(a), 9(b) and 9(c).
One can find two peaks in figure 9(a),
 one below the Fermi energy level and the other
 almost at the Fermi energy level.
The value of $S^{2}\sigma/\tau$ is maximum at 400K
 for both the peaks and is found to decrease
 with a decrease in temperature.
The same observation is seen in figures 9(b) and 9(c),
 except that there is an additional peak in the case
 of the scheme drawn with SOC,
 where the peak lies deep inside the Fermi level.

\section{Conclusions}
\label{sect-conclusion}

The first principles calculation of the half Heusler
 \ce{DyPdBi} compound is investigated and presented
 by using the full potential - linear augmented plane wave
 (FP-LAPW) method as implemented in the Wien2k code.
Different exchange correlation potentials,
 namely PBEsol with and without spin orbit coupling,
 spin polarised LSDA, and LSDA+U were employed.
The optimised structural parameter of
 $a_{opt} =12.4114 a.u.$ was obtained by using LSDA+U
 and it matches well with the experimental data.
Spin polarised calculations reveal that
 the density of states at the Fermi energy level
 is greater for spin-down states;
 it is very feeble for spin-up states.
The spin magnetic moment per unit cell ranges
 from 4.73 to 4.87 $\mu_B$ in the above study.
The density of states histogram
 drawn for PBEsol with spin orbit coupling
 reveals the splitting of dysprosium f-states
 at the Fermi energy level.
This is to be appreciated,
 as it is one of the requirements
 for being a good thermoelectric material.
There is approximately a 50 \% reduction
 in the density of states
 of \ce{Dy} f-electron states at the Fermi level
 by the inclusion of spin orbit coupling.
The inclusion of onsite coulomb interaction LSDA+U
 increases the total DoS at the Fermi energy level
 to 30 states/eV from 25 states/eV.
The band structure of spin polarised LSDA+U calculations
 both for spin-up and spin-down states augments the
 possibility of \ce{DyPdBi} to be
 a potential candidate for TE applications.
The band structure of spin-up states reveals
 the presence of a Lorentzian narrow peak
 with a sharp singularity very close to the Fermi level,
 which is a signature for a potential thermoelectric material.
Further, for spin-down states,
 the presence of localised flat bands
 that correspond to infinite effective mass
 at the valence band maximum is another key factor
 to improve the thermoelectric efficiency.

The calculated power factor 30.8 $\mu$W$cm^{-1}K^{-2}$
 using PBEsol with SOC,
 and 34.76 $\mu$W$cm^{-1}K^{-2}$ by LSDA+U,
 at Fermi energy for 300K
 are in good agreement with
 the experimentally reported values.

The above band profile investigations on \ce{DyPdBi}
 suggest that it may serve as
 a good thermoelectric material.
 Further its efficiency can be improved by doping
 which allows one to tune the band structure of \ce{DyPdBi}
 suitably in order to obtain maximum TE efficiency.

\section{Acknowledgement}

MS thanks Prof. Dr. M. Rajagopalan of the Anna University
 for his constant support and encouragement,
 and Prof. Sushil Auluck, NPL, New Delhi
 for his valuable suggestions.





\newpage


\noindent\textbf{Table 1. Structural, Electronic and
 Magnetic properties of DyPdBi}

$\,$

\noindent\begin{tabular}{ |m{4cm}|c|c|c| }
\hline
 & \multicolumn{3}{|c|}{\textbf{Spin Polarised}} \\
\cline{2-4}
 & & & \\
 & \textbf{LSDA} & \textbf{PBE} & \textbf{(LSDA+U)} \\
\textbf{Parameter} & & & \\
\hline
 & & & \\
Lattice Parameter \newline
 (a.u) \newline
 $a_{expt}$ = 12.53884 [15] &
 $a_{opt}$ = 12.3808 &
 $a_{opt}$ = 12.7472 &
 $a_{opt}$ = 12.4114 \\
 & & & \\
\hline
 & & & \\
$E_{F}$ (Ryd) & 0.534 & 0.4855 & 0.5429 \\
 & & & \\
\hline
 & & & \\
$E_{tot}$ (Ryd) & -77545.867563 &
 -77594.12155 & -77545.76333 \\
 & & & \\
\hline
 & & & \\
B (GPa) & 115.4130 & 70.2516 & 88.6747 \\
 & & & \\
\hline
 & & & \\
$N(E_{F})$$\uparrow$/Ryd & 3.02 & 3.71 & 4.64 \\
 & & & \\
\hline
 & & & \\
$N(E_{F})$$\downarrow$/Ryd & 482.56 & 178.30 & 584.80 \\
 & & & \\
\hline
 & & & \\
Spin Magnetic \newline
 moment per \newline
 unit cell ($\mu_{B}$) &
 4.87 & 4.83 & 4.73 \\
 & & & \\
\hline
\end{tabular}

\newpage

\begin{raggedright}
\noindent\textbf{Table 2. Transport Properties of DyPdBi
  at 300 K for Fermi \\ Energy}
\end{raggedright}

$\,$

\noindent\begin{tabular}{ |m{3cm}|m{2cm}|m{2cm}|m{2cm}|m{2.5cm}|m{2cm}| }
\hline
 & & & & & \\
Method &
Seebeck \newline
 Coefficient \newline
 (S) \newline
 ($\mu$V/K) &
Resistivity \newline
 ($\rho$) \newline
 ($\mu\Omega$-cm) &
$\sigma/\tau$ \newline
 $10^{19}$($\Omega^{-1}$ \newline
 $m^{-1}s^{-1}$) &
$S^2\sigma/\tau$ \newline
 $10^{14}$($\mu$W \newline
 $cm^{-1}K^{-2}s^{-1}$) &
Power \newline
 Factor \newline
 ($\mu$W \newline
 $cm^{-1}K^{-2}$)) \\
 & & & & & \\
\hline
 & & & & & \\
Without SOC & -107.1 & 381.381 & 3.28 & 37.6 & 30.09 \\
 & & & & & \\
\hline
 & & & & & \\
With SOC & -138 & 618.465 & 1.04 & 19.8 & 30.81 \\
 & & & & & \\
\hline
 & & & & & \\
(LSDA+U) for \newline
 spin down & -113 & 370.547 & 3.38 & 43.5 & 34.76 \\
 & & & & & \\
\hline
 & & & & & \\
(LSDA+U) for \newline
 spin up & 8.61 & 55.306 & 22.6 & 1.674 & 1.339 \\
 & & & & & \\
\hline
\end{tabular}

\newpage



\begin{figure}
\centering
\begin{subfigure}[b]{0.8\textwidth}
\includegraphics[width=1\linewidth]{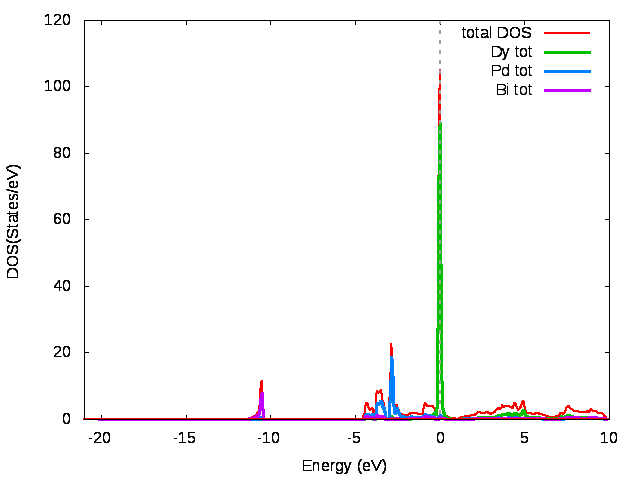}
\caption{}
\label{fig1a}
\end{subfigure}
\begin{subfigure}[b]{0.8\textwidth}
\includegraphics[width=1\linewidth]{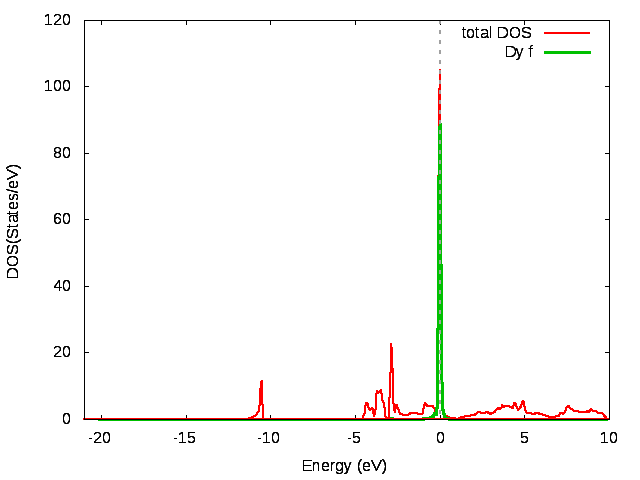}
\caption{}
\label{fig1b}
\end{subfigure}
\caption*{\textbf{Figure 1. (a) Total and (b) partial DoS
 histograms of DyPdBi \underline{without SOC}.}}
\end{figure}

\begin{figure}
\centering
\begin{subfigure}[b]{0.8\textwidth}
\includegraphics[width=1\linewidth]{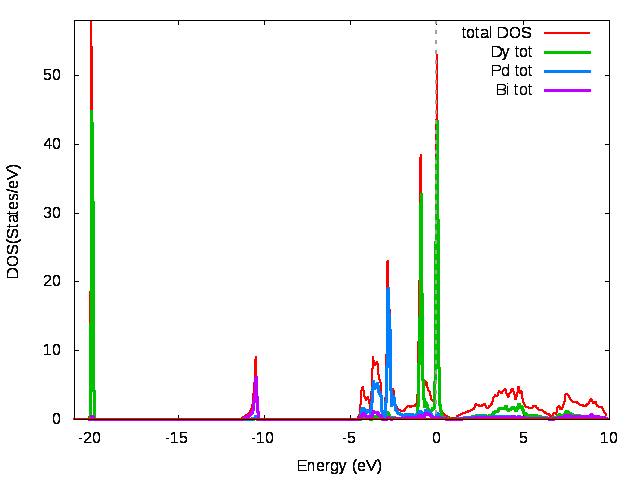}
\caption{}
\label{fig2a}
\end{subfigure}
\begin{subfigure}[b]{0.8\textwidth}
\includegraphics[width=1\linewidth]{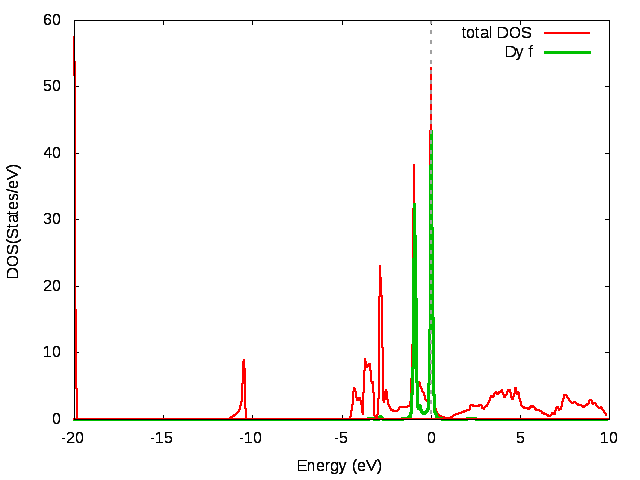}
\caption{}
\label{fig2b}
\end{subfigure}
\caption*{\textbf{Figure 2. (a) Total and (b) partial DoS
 histograms of DyPdBi \underline{with SOC}.}}
\end{figure}

\begin{figure}
\centering
\begin{subfigure}[b]{0.8\textwidth}
\includegraphics[width=1\linewidth]{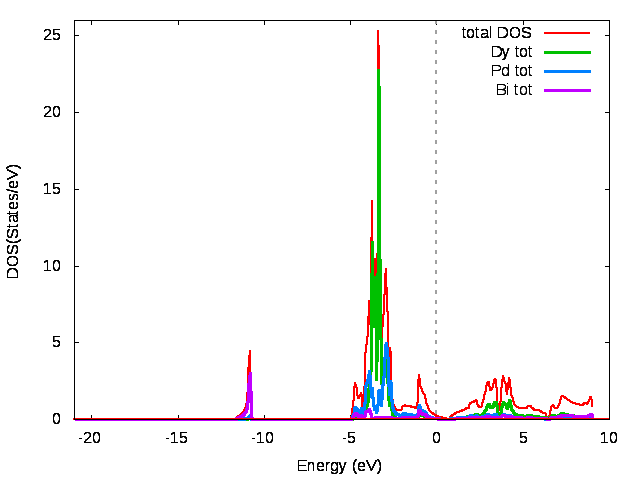}
\caption{spin up}
\label{fig3a}
\end{subfigure}
\begin{subfigure}[b]{0.8\textwidth}
\includegraphics[width=1\linewidth]{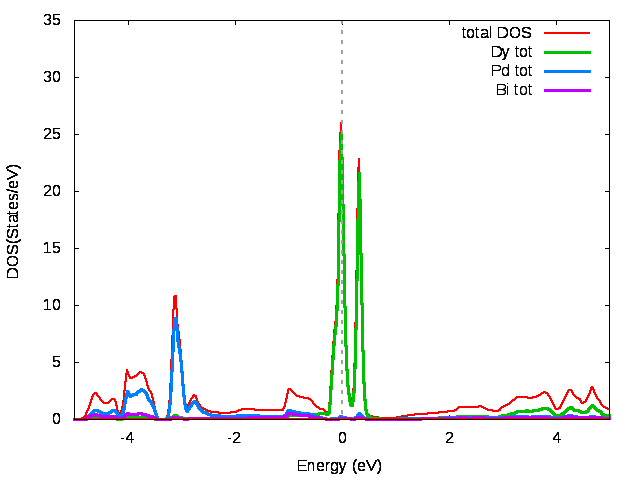}
\caption{spin down}
\label{fig3b}
\end{subfigure}
\caption*{\textbf{Figure 3. DoS histograms for
 (a) \underline{spin up} and (b) \underline{spin down}
 states of DyPdBi by LSDA.}}
\end{figure}

\begin{figure}
\centering
\begin{subfigure}[b]{0.8\textwidth}
\includegraphics[width=1\linewidth]{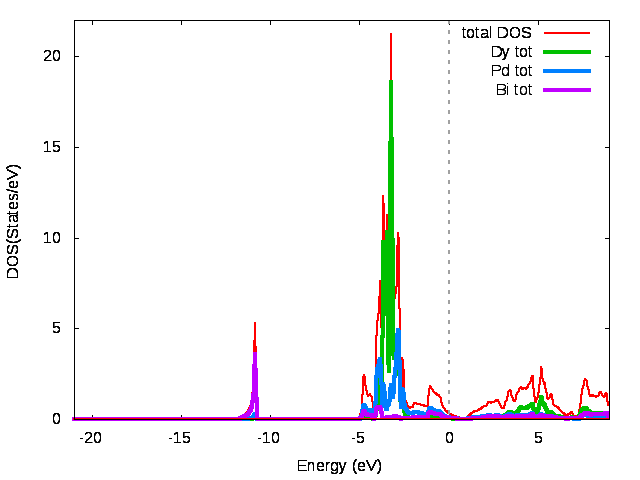}
\caption{spin up}
\label{fig4a}
\end{subfigure}
\begin{subfigure}[b]{0.8\textwidth}
\includegraphics[width=1\linewidth]{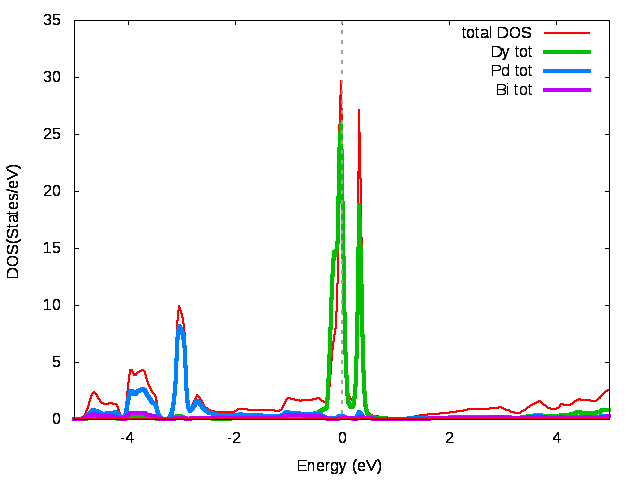}
\caption{spin down}
\label{fig4b}
\end{subfigure}
\caption*{\textbf{Figure 4. DoS histograms for
 (a) \underline{spin up} and (b) \underline{spin down}
 states of DyPdBi by (LSDA+U).}}
\end{figure}

\begin{figure}
\centering
\begin{subfigure}[b]{0.45\textwidth}
\includegraphics[width=1\linewidth]{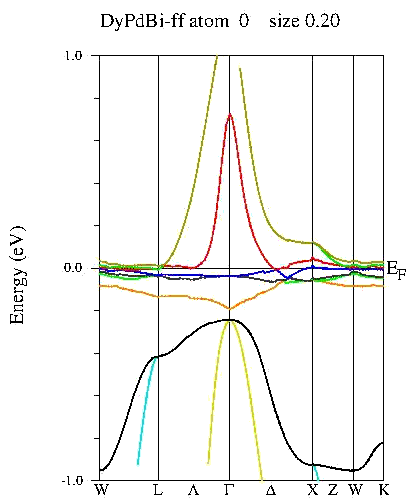}
\caption{without SOC}
\label{fig5a}
\end{subfigure}
\begin{subfigure}[b]{0.45\textwidth}
\includegraphics[width=1\linewidth]{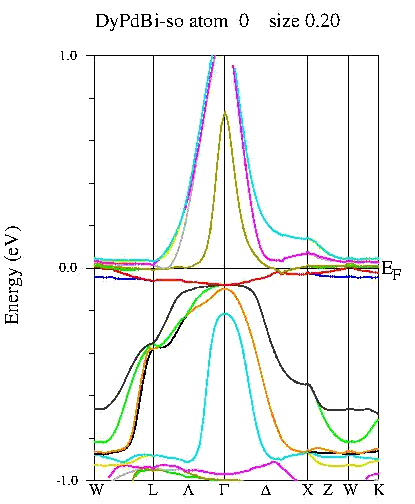}
\caption{with SOC}
\label{fig5b}
\end{subfigure}
\caption*{\textbf{Figure 5. Band structure of DyPdBi drawn
 (a) \underline{without SOC} and (b) \underline{with SOC}.}}
\end{figure}

\begin{figure}
\centering
\begin{subfigure}[b]{0.45\textwidth}
\includegraphics[width=1\linewidth]{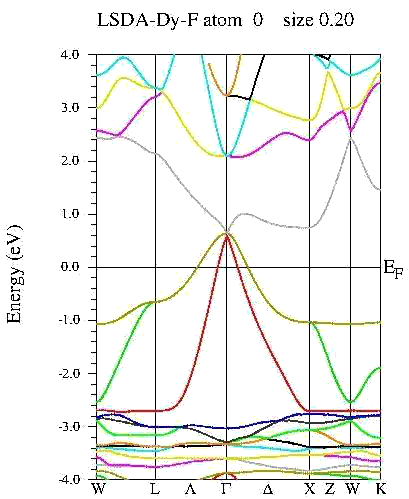}
\caption{LSDA}
\label{fig6a}
\end{subfigure}
\begin{subfigure}[b]{0.45\textwidth}
\includegraphics[width=1\linewidth]{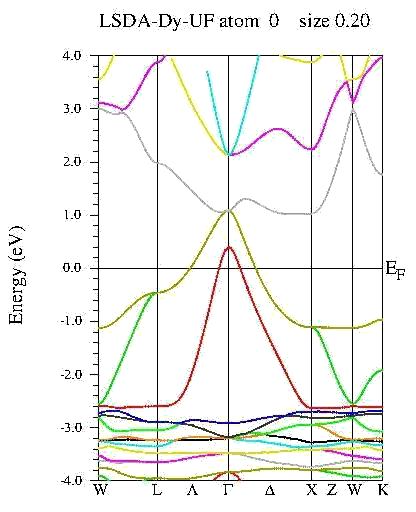}
\caption{(LSDA+U)}
\label{fig6b}
\end{subfigure}
\caption*{\textbf{Figure 6. Band structure of
 spin-up states of DyPdBi
 (a) \underline{for LSDA} and (b) \underline{for (LSDA+U)}.}}
\end{figure}

\begin{figure}
\centering
\begin{subfigure}[b]{0.45\textwidth}
\includegraphics[width=1\linewidth]{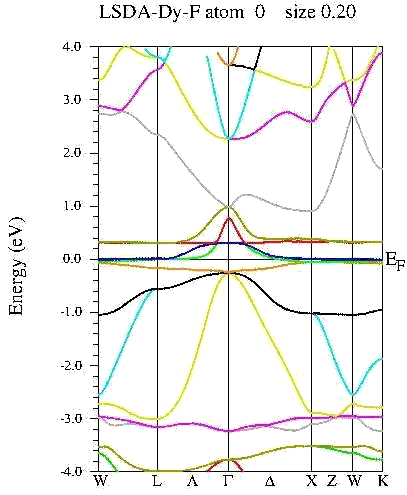}
\caption{LSDA}
\label{fig7a}
\end{subfigure}
\begin{subfigure}[b]{0.45\textwidth}
\includegraphics[width=1\linewidth]{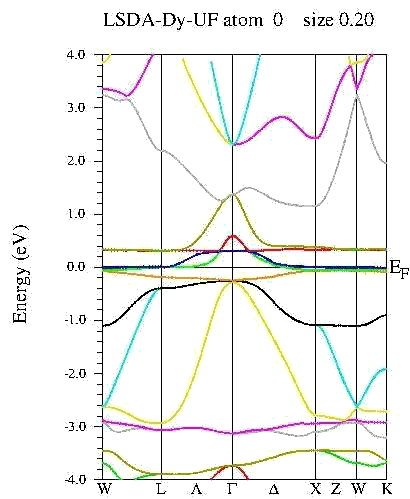}
\caption{(LSDA+U)}
\label{fig7b}
\end{subfigure}
\caption*{\textbf{Figure 7. Band structure of
 spin-down states of DyPdBi
 (a) \underline{for LSDA} and (b) \underline{for (LSDA+U)}.}}
\end{figure}

\begin{figure}
\centering

\begin{subfigure}[b]{0.55\textwidth}
\includegraphics[width=1\linewidth]{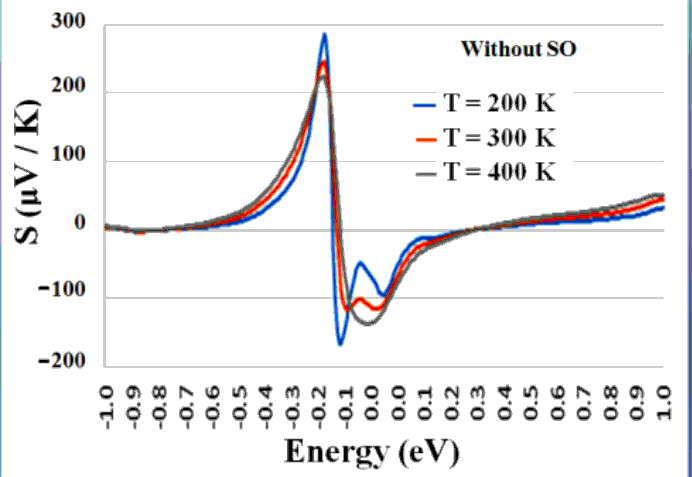}
\caption{without SOC}
\label{fig8a}
\end{subfigure}

\begin{subfigure}[b]{0.55\textwidth}
\includegraphics[width=1\linewidth]{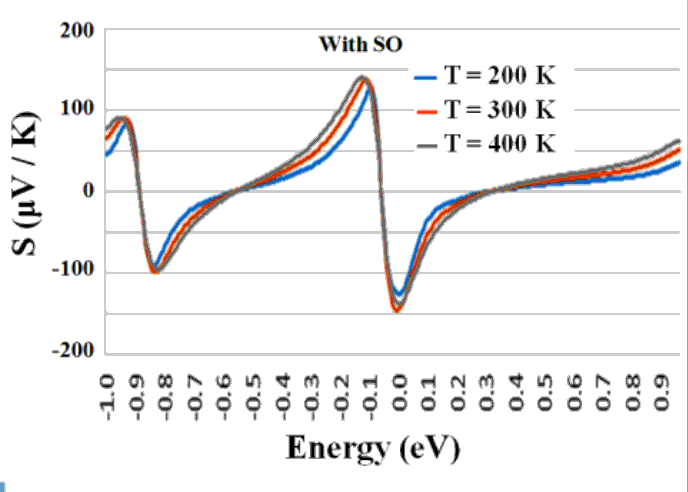}
\caption{with SOC}
\label{fig8b}
\end{subfigure}

\begin{subfigure}[b]{0.55\textwidth}
\includegraphics[width=1\linewidth]{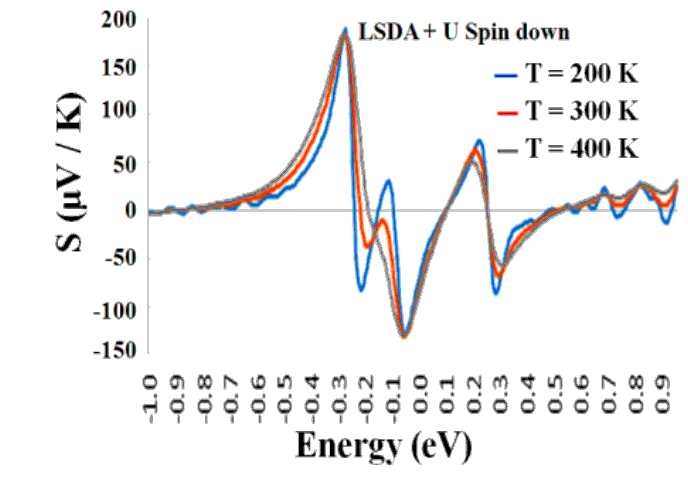}
\caption{LSDA+U Spindown}
\label{fig8c}
\end{subfigure}

\caption*{\textbf{Figure 8.
 DyPdBi Seebeck Coefficient vs total energy.}}
\end{figure}

\begin{figure}
\centering

\begin{subfigure}[b]{0.55\textwidth}
\includegraphics[width=1\linewidth]{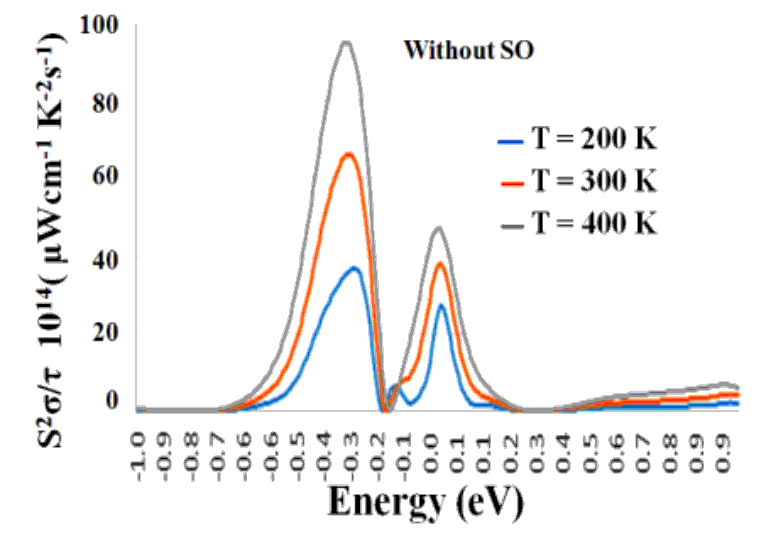}
\caption{without SOC}
\label{fig9a}
\end{subfigure}

\begin{subfigure}[b]{0.55\textwidth}
\includegraphics[width=1\linewidth]{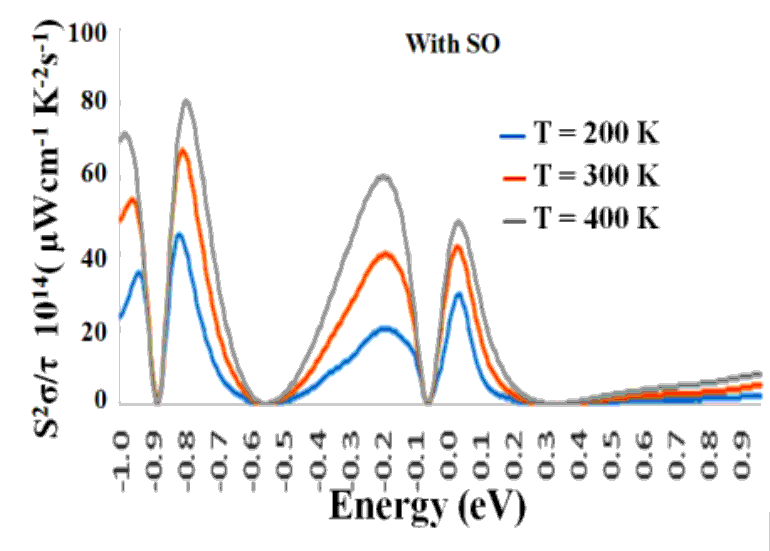}
\caption{without SOC}
\label{fig9b}
\end{subfigure}

\begin{subfigure}[b]{0.55\textwidth}
\includegraphics[width=1\linewidth]{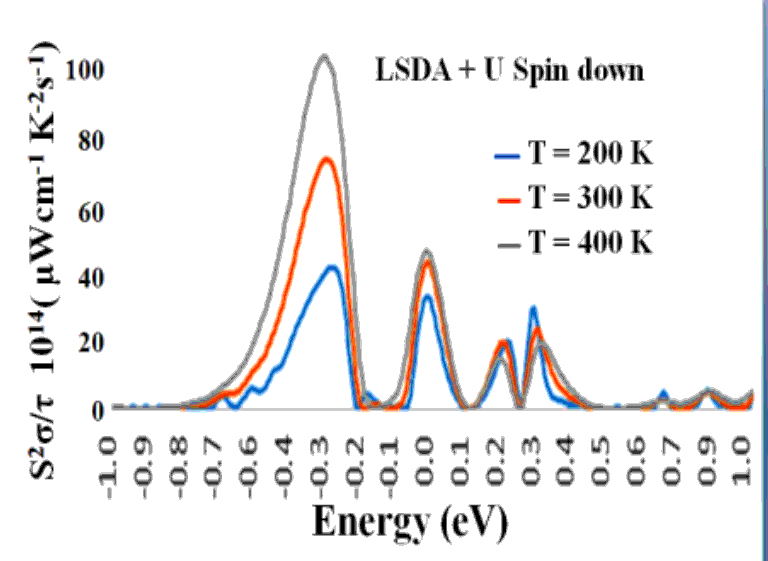}
\caption{LSDA+U Spindown}
\label{fig9c}
\end{subfigure}

\caption*{\textbf{Figure 9.
 DyPdBi $S^2\sigma/\tau$ vs total energy.}}
\end{figure}

\end{document}